\documentclass[preprint,12pt]{elsarticle}

\usepackage{amssymb}
\usepackage{amsmath}
\usepackage{mathtools}
\usepackage{listings}
\usepackage{subcaption}

\usepackage[draft]{minted}

\newcounter{bla}

\newcommand{\dpv}{$\partial PV$}
\newcommand{\citeasnoun}[1]{Ref.~\citenum{#1}}

\journal{Computer Physics Communications}

\begin{document}

\begin{frontmatter}



\title{\dpv: An End-to-End Differentiable Solar-Cell Simulator}


\author[a]{Sean Mann}
\author[a]{Eric Fadel}
\author[b]{Samuel S. Schoenholz}
\author[b]{Ekin D. Cubuk}
\author[a]{Steven G. Johnson}
\author[a]{Giuseppe Romano}
\cortext[author] {Corresponding authors.\\\textit{E-mail address:}seanmann@mit.edu, romanog@mit.edu}
\address[a]{Massachusetts Institute of Technology, 77 Massachusetts Avenue, Cambridge, Massachusetts 02139, USA}
\address[b]{Google Brain, Google Inc., Mountain View, California 94043, USA}

\begin{abstract}
We introduce \dpv, an end-to-end differentiable photovoltaic (PV) cell simulator based on the drift-diffusion model and Beer--Lambert law for optical absorption. \dpv is programmed in Python using JAX, an automatic differentiation (AD) library for scientific computing. Using AD coupled with the implicit function theorem, \dpv computes the power conversion efficiency (PCE) of an input PV design as well as the derivative of the PCE with respect to any input parameters, all within comparable time of solving the forward problem. We show an example of perovskite solar-cell optimization and multi-parameter discovery, and compare results with random search and finite differences. The simulator can be integrated with optimization algorithms and neural networks, opening up possibilities for data-efficient optimization and parameter discovery. 

\end{abstract}


\end{frontmatter}

{\bf PROGRAM SUMMARY}

\begin{small}
\noindent
{\em Program Title: \dpv}                                          \\
{\em CPC Library link to program files:} (to be added by Technical Editor) \\
{\em Developer's repository link:} https://github.com/romanodev/deltapv.git \\
{\em Code Ocean capsule:} (to be added by Technical Editor)\\
{\em Licensing provisions:} MIT  \\
{\em Programming language: Python}                                   \\
{\em Nature of problem:} Photovoltaic cell optimization has been traditionally difficult due to the lack of gradients from numerical drift-diffusion solvers. This results in the need to treat the problem as a case of black-box optimization, which incurs high computational costs and low data efficiency. \\
{\em Solution method:} An end-to-end differentiable photovoltaic simulator via the drift-diffusion model was developed using JAX, a growing scientific computation and automatic-differentiation library. To enhance computational speed, the implicit function theorem was used to bypass the need for directly differentiating through iterative solvers. \\

\end{small}

\section{Introduction}
\label{intro}

Numerical modelling of solar cells has become an essential tool of photovoltaic research. A number of packages have been extensively developed, and are freely available, such as PC-1D \cite{pc1d}, SCAPS \cite{scaps}, wxAMPS \cite{wxamps} or SESAME \cite{sesame}. Such simulators have brought significant advances in studying solar cells, and have been applied to common systems including CdS-CdTe \cite{cdscdte} and perovskite cells \cite{perov}. These computational tools also allow the study of important physical phenomena such as the effects of grain boundaries \cite{grain} and defects \cite{defects}, that impact the performance of solar cells. Although solar cell modelling is quite well developed, the optimization of solar cells is much more challenging, particularly in the case of optimizing many design variables jointly. This is in part due to the unavailability of derivatives, which are key to efficient high-dimensional optimization.

In this work we present \dpv, a 1D simulation tool for PV cells which solves the drift-diffusion equations using the JAX automatic differentiation (AD) package \cite{jax2018github}. \dpv is able to compute not only the efficiency of the solar cell but also its derivative with respect to any material property set by the user. Thus, this new computational tool enables extensive, efficient materials optimization for PV cells, and can be used in conjunction with standard optimization methods, or machine learning algorithms. By computing derivatives with AD, we avoid numerical and scaling issues associated with the finite-difference technique, while adding complexity comparable to that of the original problem~\cite{baydin2018automatic}. 
The rise of this line of research has been enabled by the development of growing AD-tools. Notable examples include JAX~\cite{jax2018github}, where AD is based on tracing, and Zygote~\cite{rackauckas2020generalized}, built on source-to-source transformation. Based on these tools, several AD-enhanced simulators have been released, including for molecular dynamics~\cite{schoenholz2020jax,eastman2017openmm}, fluid dynamics~\cite{kochkov2021machine}, kinetics~\cite{goodrich2021designing}, optics~\cite{oskooi2010meep} and general purpose solvers~\cite{hu2019difftaichi}. 
\dpv, therefore, complements the existing offer of AD-based solvers and, relying on the composability of these tools, potentially enables end-to-end differentiability of more advanced multiphysics simulations where the DD model is coupled with full Maxwell equations. 

The manuscript is organized as follows. We first provide a brief overview of the drift-diffusion model and its parameters. Second, we detail several technical aspects of the gradient calculations. In the third section, we describe the software architecture and its API. Then, we show an example optimization of a perovskite solar cell. Finally, we use \dpv for material parameter discovery. \dpv is a Python package freely distributed under the ``MIT'' license, also known as the ``Expat license.'' This permissive licence will allow other entities to incorporate \dpv in their software with minimal restrictions, thus further accelerating the integration of our tool in multiphysics simulators.”

\section{Model}
\label{model}

Solar cells are commonly modelled by the drift-diffusion (DD) model, a set of coupled nonlinear differential equation describing charge dynamics under illumination. Although this approach suffers from several limitations, the main one being its simplified treatment of band diagrams, it offers a compelling trade-off between accuracy and efficiency. A comprehensive review can be found in \citeasnoun{nelson2003physics}. As detailed in~\ref{physics}, the DD model is a nonlinear relationship between the electrostatic potential $\phi(\mathbf{r})$ and the electrochemical potentials for electrons ($\phi_n(\mathbf{r})$) and holes ($\phi_p(\mathbf{r})$), which, after discretization, takes the form
\begin{equation}\label{linear}
\mathbf{f}(\mathbf{u},\mathbf{p}) = 0;
\end{equation}
where the unknown $\mathbf{u} = \mathrm{vec}(\boldsymbol{\phi},\boldsymbol{\phi}_n,\boldsymbol{\phi}_p)$ is a vector of the discretized potentials. The term $\mathbf{p}$ includes $M$ material parameters, the reverse voltage~$V$, and the illumination~$S$ (used to calculate the carrier generation density via the Beer--Lambert law; see \ref{optics}). The boundary conditions for both the equilibrium case (e.g.~$S=0$) and general case are reported in \ref{boundary}. The function $f$ appearing in Eq.~\ref{linear} represents a set of coupled nonlinear differential equations, and is solved using the Newton--Raphson method: starting from an initial guess, $\mathbf{u}^{0}$, successive steps are the solution of the linear system:
\begin{equation}\label{nn}
 \mathbf{f}_{\mathbf{u}} (\mathbf{u}^{(i)}) \Delta \mathbf{u}^{(i)} = -\mathbf{f}(\mathbf{u}^{(i)}), 
\end{equation}
where $\mathbf{f}_{\mathbf{u}}^{-1}$ is the Jacobian of $\mathbf{f}$. Equation~\ref{nn} is stopped upon convergence, i.e. when $|\Delta \mathbf{u}^{(i)}| < \epsilon$. Note that we use subscripts to refer to partial derivatives, except when it is clear from the context that it is a vector index. The Newton method displays quadratic convergence only when the trial function is sufficiently close to the root~\cite{newton}, and is not numerically robust when used on its own in practice. Furthermore, the DD model contains exponentials of the potentials, often resulting in convergence stagnation. To mitigate these issues, we employ a modified version of the Newton algorithm, as described in~\ref{newton}. The DD itself is discretized using the Scharfetter--Gummel scheme, which is based on finite differences (for details see~\ref{numerical}). Lastly, the linear system in Eq.~\ref{nn} is solved with the preconditioned generalized minimal residual method (GMRES), which was recently included in JAX. Since this method is \textit{matrix-free}, it allows us to exploit the sparsity of our problem by directly providing the linear operator representing the product of the Jacobian and an arbitrary vector. For these problems, a preconditioner can dramatically speed up convergence; we use an $ILU(0)$ preconditioner, as detailed in in~\ref{solver}.  In order to compute the power conversion efficiency (PCE) of a PV cell, the DD equations must be solved for multiple reverse bias voltages from zero up to the open-circuit voltage $V_{oc}$ where the current through the cell reaches zero. The efficiency is computed as
\begin{equation}
  \eta = \frac{P_{\mathrm{MPP}}}{P_{\mathrm{in}}}
\end{equation}
where $P_{\mathrm{in}}$ is the power of the incoming radiation, and $P_{\mathrm{MPP}}$ is the maximal power point, i.e.~the maximum electrical power $P = I(V) \cdot V$, $I$ being the electrical current. The device optimization, therefore, requires the gradient of the output power with respect to $V$ as well as to the device parameters $\mathbf{p}$. In the case of maximizing the efficiency of a cell over material parameters, the bias voltage $V$ can be treated simply as an additional optimization variable. For convenience, let us define the vector $\tilde{\mathbf{p}} = [V,\mathbf{p}^\top]^\top$; the gradient of the power output with respect to $\mathbf{\tilde{p}}$ is

\begin{equation}\label{grad}
    \frac{d P_{\mathrm{out}}}{d\mathbf{\tilde{p}}} = \frac{d \left(I V\right)}{d\mathbf{\tilde{p}}} = V\frac{d I}{d \mathbf{\tilde{p}}} + I \mathbf{e}_1^T
\end{equation}
where $\mathbf{e}_1$ refers to the unit vector corresponding to the first component of $\tilde{\mathbf{p}}$ (i.e.~$V$). Note that $I$ depends on $\mathbf{\tilde{p}}$ both directly and via the solutions $\mathbf{u}$:

\begin{equation}\label{cur}
\frac{d I}{d\mathbf{\tilde{p}}} = I_{\mathbf{u}}\mathbf{u}_{\mathbf{\tilde{p}}} + I_{\mathbf{\tilde{p}}},
\end{equation}
where we used subscripts to indicate partial derivatives. The terms $I_{\mathbf{u}}$ and $I_{\tilde{\mathbf{p}}}$ in Eq.~\ref{cur} are readily available using AD; however, differentiating $\mathbf{u}$ with respect to $\tilde{\mathbf{p}}$ requires care. Since \dpv is built in JAX, which supports taking arbitrary gradients and Jacobians of functions, it is technically possible to directly differentiate the PDE solution using AD. In practice, the large number of iterations, including those from the Newton--Raphson method, need to be unrolled and differentiated through the chain rule, which leads to significant computation time and memory usage. Fortunately, there is a simple way to bypass differentiating through the solver: implicitly differentiating the PDE system using the implicit function theorem (IFT)~\cite{ift,margossian2019review}. A brief overview of the IFT applied to our case is given in~\ref{ifta}.

\section{Software Architecture}
\label{arch}

\subsection{Simulation object}

\dpv revolves around the \verb|PVDesign| object, which contains all design parameters of a candidate PV cell. JAX follows a functional programming paradigm to facilitate defining gradients of actions performed on immutable, stateless objects. Hence, \dpv defines a series of operations on an immutable \verb|PVDesign|, each returning a new object.

We illustrate the usage of \dpv with a simple example. First, we import necessary packages:

\begin{minted}[breaklines]{python}
  import deltapv as dpv
  from jax import numpy as jnp, grad
\end{minted}
Note that we use JAX's numpy, which has virtually the same API as the well-known numpy package \cite{numpy} with the exception of arrays being immutable. In this example, we work with a p--n homojunction. First, we define a custom material as follows:

\begin{minted}[breaklines]{python}
material = dpv.create_material(Chi=3.9,
                               Eg=1.5,
                               eps=9.4,
                               Nc=8e17,
                               Nv=1.8e19,
                               mn=100,
                               mp=100,
                               tn=1e-8,
                               tp=1e-8,
                               A=2e4)
\end{minted}

\dpv is also equipped with a rudimentary materials library, including several common semiconductor materials. To load silicon, for example, one could instead use
\begin{minted}[breaklines]{python}
material = dpv.load_material("Si")
\end{minted}

\dpv has a simple API that, for most purposes, enables defining a \verb|PVDesign| object in one line. To define a p-n junction of thickness $2 \times 10^{-4} ~ \text{cm}$, where the junction is in the middle, and discretized on 500 uniform grid points, we can use

\begin{minted}[breaklines]{python}
des = dpv.make_design(n_points=500,
                      Ls=[1e-4, 1e-4],
                      mats=material,
                      Ns=[1e17, -1e17],
                      Snl=1e7,
                      Snr=0,
                      Spl=0,
                      Spr=1e7)
\end{minted}
which creates such a junction with donor density $1 \times 10^{17} ~ \text{cm}^{-3}$ on the left and an equal acceptor density on the right.
The surface recombination velocities are as specified, in $\text{cm/s}$.

\begin{figure}
  
  \begin{subfigure}[t]{.5\textwidth}
    \centering
    \includegraphics[width=\linewidth]{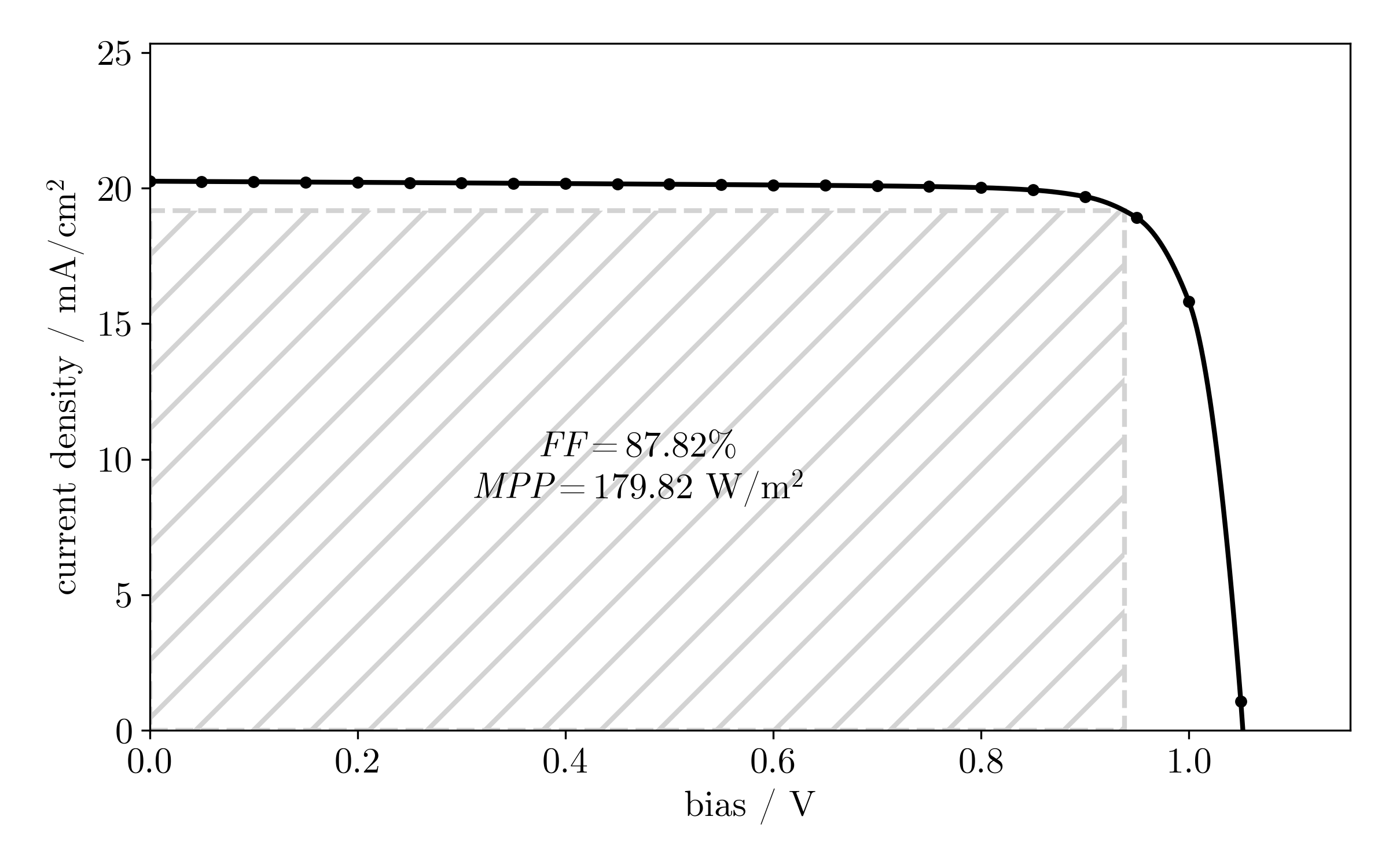}
    \caption{}
  \end{subfigure}
  \hfill
  \begin{subfigure}[t]{.5\textwidth}
    \centering
    \includegraphics[width=\linewidth]{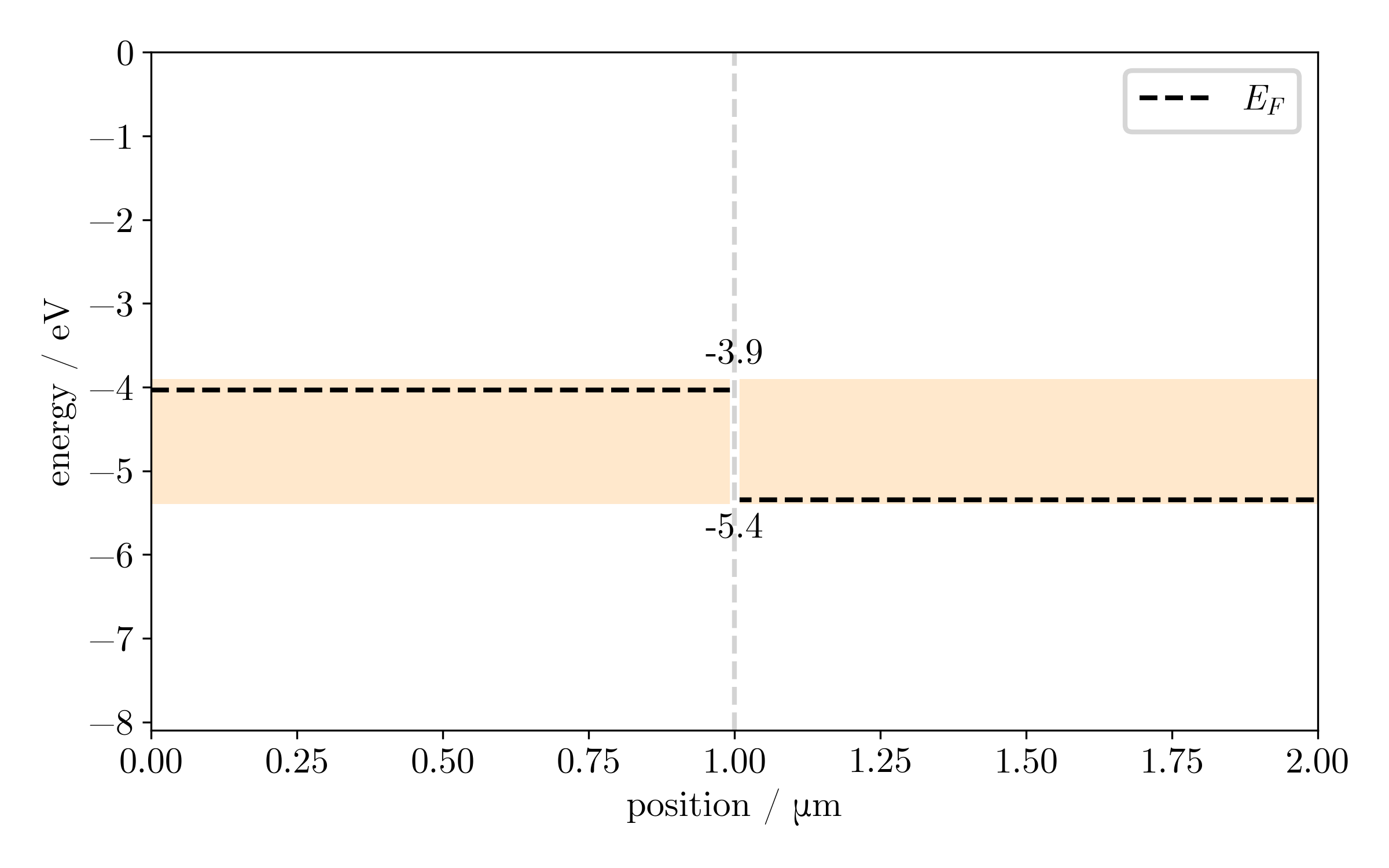}
    \caption{}
  \end{subfigure}

  \medskip

  \begin{subfigure}[t]{.5\textwidth}
    \centering
    \includegraphics[width=\linewidth]{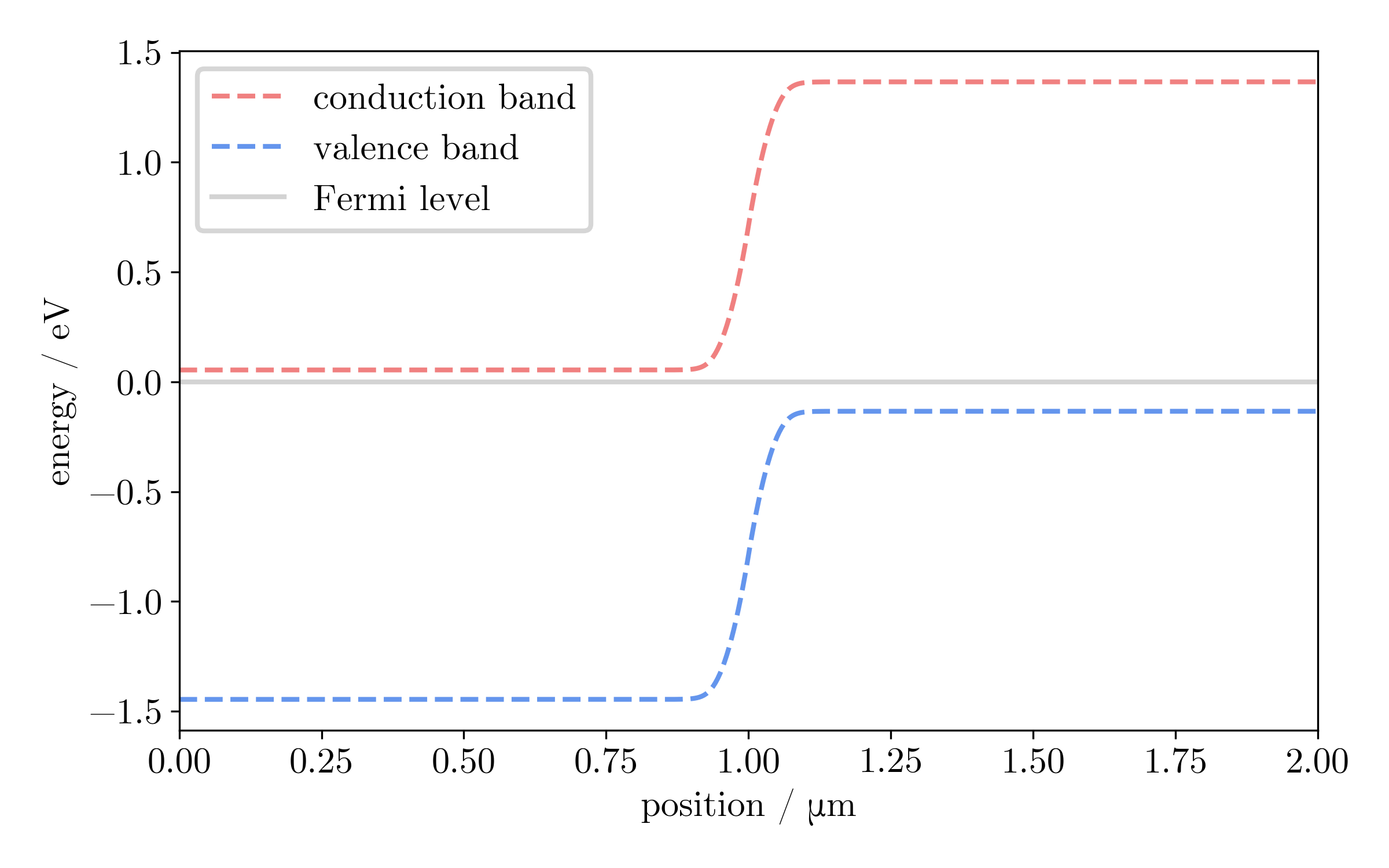}
    \caption{}
  \end{subfigure}
  \hfill
  \begin{subfigure}[t]{.5\textwidth}
    \centering
    \includegraphics[width=\linewidth]{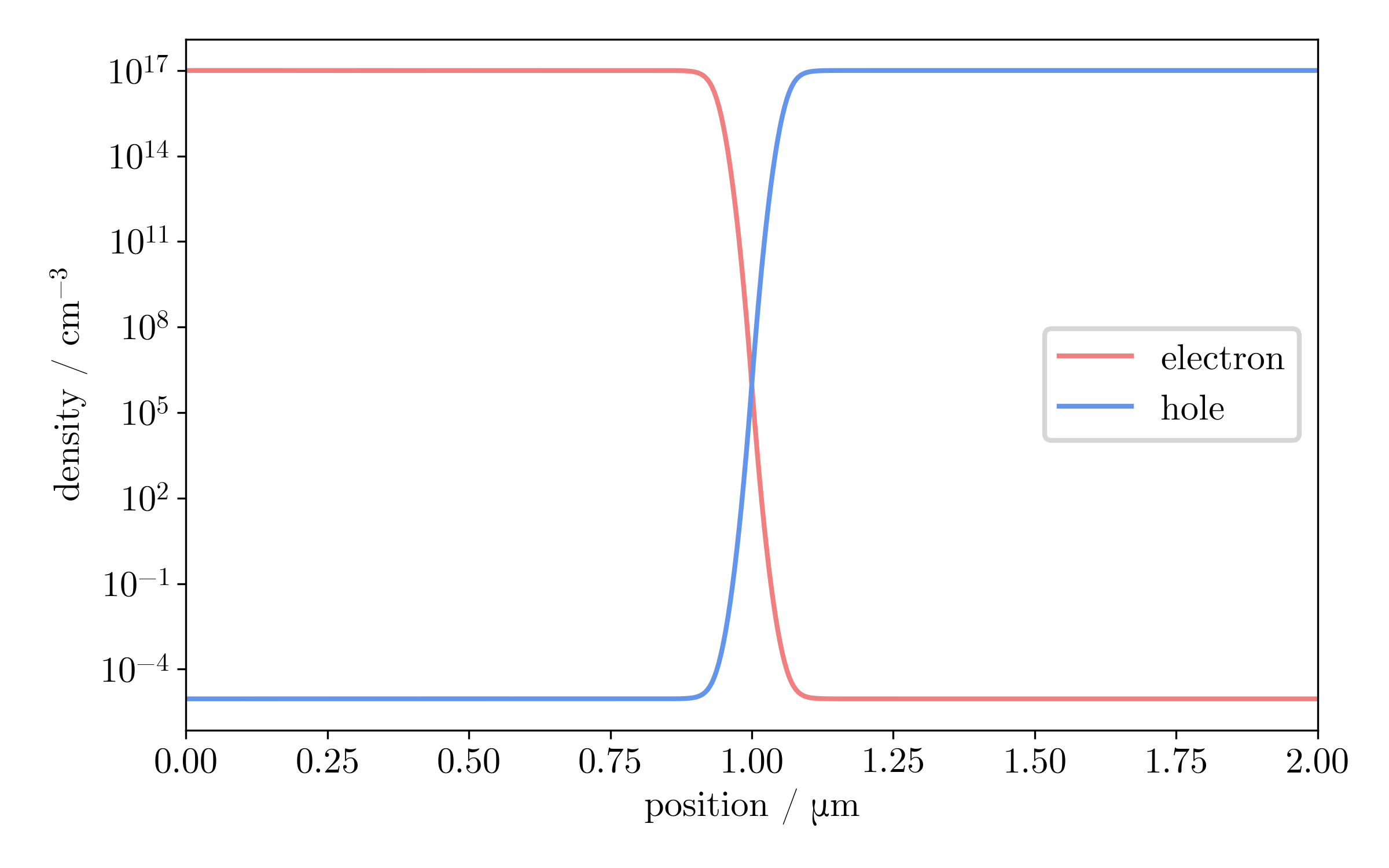}
    \caption{}
  \end{subfigure}
  \caption{The case of a p-n homojunction. (a) $IV$ curve, (b) isolated energy levels, (c) equilibrium band diagram, (d) equilibrium carrier densities.}
  \label{fig:example}
\end{figure}

The simulation can be started with
\begin{minted}[breaklines]{python}
results = dpv.simulate(des)
\end{minted}
where the equilibrium case is first computed. The solver then applies an increasing reverse bias until $V_{oc}$ is reached. The light source used defaults to the solar spectrum, and can be specified with a \verb|LightSource| object, which is passed into \verb|simulate|, if necessary. The final result, including the PCE and the $IV$ curve, is then stored in a dictionary. \dpv determines the efficiency of this cell to be $19.98\%$. Several plotting functions are provided, which can be generated via

\begin{minted}[breaklines]{python}
dpv.plot_iv_curve(*results["iv"])
dpv.plot_bars(des)
dpv.plot_band_diagram(des, results["eq"], eq=True)
dpv.plot_charge(des, results["eq"])
\end{minted}
giving the plots in Figure \ref{fig:example}.

The $IV$ curve and band diagram for this example have been checked against SESAME \cite{sesame}, with an error within 0.1$\%$. Note that SESAME does not provide an optical model, and thus the generation density $G(x)$ computed by \dpv was extracted and directly passed into the SESAME simulation.

Taking the gradients is as simple as treating \verb|simulate| as any other differentiable function in JAX's API. To get the partial derivatives of the PCE with respect to all parameters of the \verb|PVDesign|, one could use

\begin{minted}[breaklines]{python}
  f = lambda des: dpv.simulate(des)["eff"]
  dfddes = grad(f)(des)
\end{minted}
which gives an object, of the same structure as a \verb|PVDesign|, but where every parameter is replaced with the \textit{derivative} of the PCE with respect to the parameter itself. Due to the availability of AD, the \dpv simulation procedure can be composed with other calculations and differentiated through directly. This means that a user may define a wrapper function around it with a small number of arguments, and only compute the derivatives with respect to these higher-level, composed parameters. For example, for a single p-n homojunction with uniform material, the problem can be parameterized with a single set of material parameters, with no dependence on $x$. This may be more useful in many cases when a parameter cannot be continuously varied over space.

\section{Examples}
\label{example}

\subsection{Sensitivity Analysis}

In this section we illustrate an example for obtaining the overall sensitivity of the efficiency with respect to a given set of material parameters. This task entails the identification of a sensitivity measure that depends on the whole parameter space. To this end, several approaches have been developed, notably the Derivative-based Global Sensitivity Measure (DGSM)~\cite{SOBOL20093009}, given by
\begin{equation}\label{sobol}
    S_i = \frac{1}{D \pi^2}\int_{H^n} \left[\frac{\partial g(\mathbf{x})}{\partial x_i}\right]^2 d\mathbf{x},
\end{equation}
where $D$ is the variance of $g(\mathbf{x})$, and $H^n$ is the hypercube defining the parameter space. In typical implementations, the derivative in Eq.~\ref{sobol} is performed using finite differences, requiring an elevated number of function evaluations. Thanks to AD, we can now compute it along the efficiency calculations. In this example, we obtain the DGSM of the efficiency for the structure presented in the previous section, with the doping kept fixed. In this example, for illustration purposes, we focus on the space spanned by the small subset of material parameters $\mu_n,\mu_p,\log \tau_n$ and $\log \tau_p$. The high-dimensional integral of Eq.~\ref{sobol} is performed using Sobol sequences; the convergence error, defined as $\epsilon_i(m) = \left(S_i(2^{m+1})-S_i(2^{m})\right)/S_i(2^{m})$, is kept below $1\%$. Note that $S_i(k)$ is the DGSM of the parameter $i$ after $k$ function evaluations. The sequences are obtained from \verb|SAliB|~\cite{herman2017salib}. The sensitivities of the material parameters as well their convergence (roughly 10 thousand evaluations needed) are reported in Fig.~\ref{sobol}; for this configuration and chosen parameters set, our calculations identify the hole mobility as the parameter having the largest overall influence on the efficiency.  

\begin{figure}
\centering
    \includegraphics[width=0.75\linewidth]{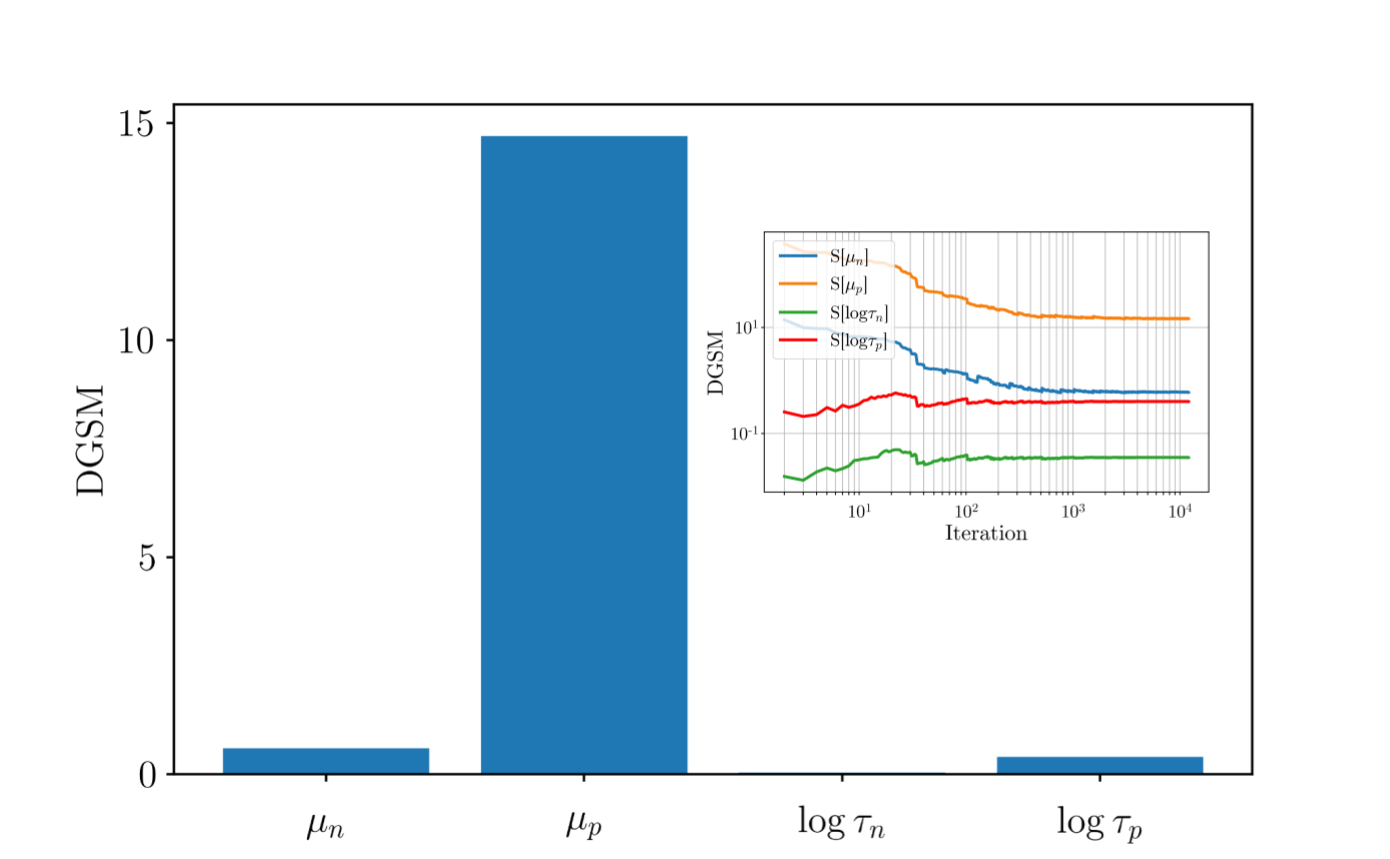}
    \caption{Global sensitivity analysis computed by~\ref{sobol}. In the inset, the convergence for each parameter is shown.}
    \label{fig:sobol}
\end{figure}

\subsection{Optimizing a p--i--n Perovskite Solar Cell}
 Traditionally, solar cell optimization has been done with a variety of gradient-free black box optimization techniques, such as Particle Swarm and Genetic Algorithm \cite{pvoptim}. Where the solar cell simulator is treated as a black box without any additional information, optimization becomes a data-intensive task. In particular, it is often intractable to simultaneously optimize more than several parameters in conjunction. Taking brute-force optimization as an example, this challenge is owing to the number of samples required to comprehensively search a set of box-constrained parameters growing exponentially in the number of parameters. Where gradients are available, however, optimization becomes a much easier task, and this becomes starker as the number of parameters grows. An arsenal of well-developed nonlinear optimization algorithms become available with the introduction of analytical gradients.

\begin{figure}
\centering
 
  \begin{subfigure}[t]{0.8\textwidth}
    \centering
    \includegraphics[width=\linewidth]{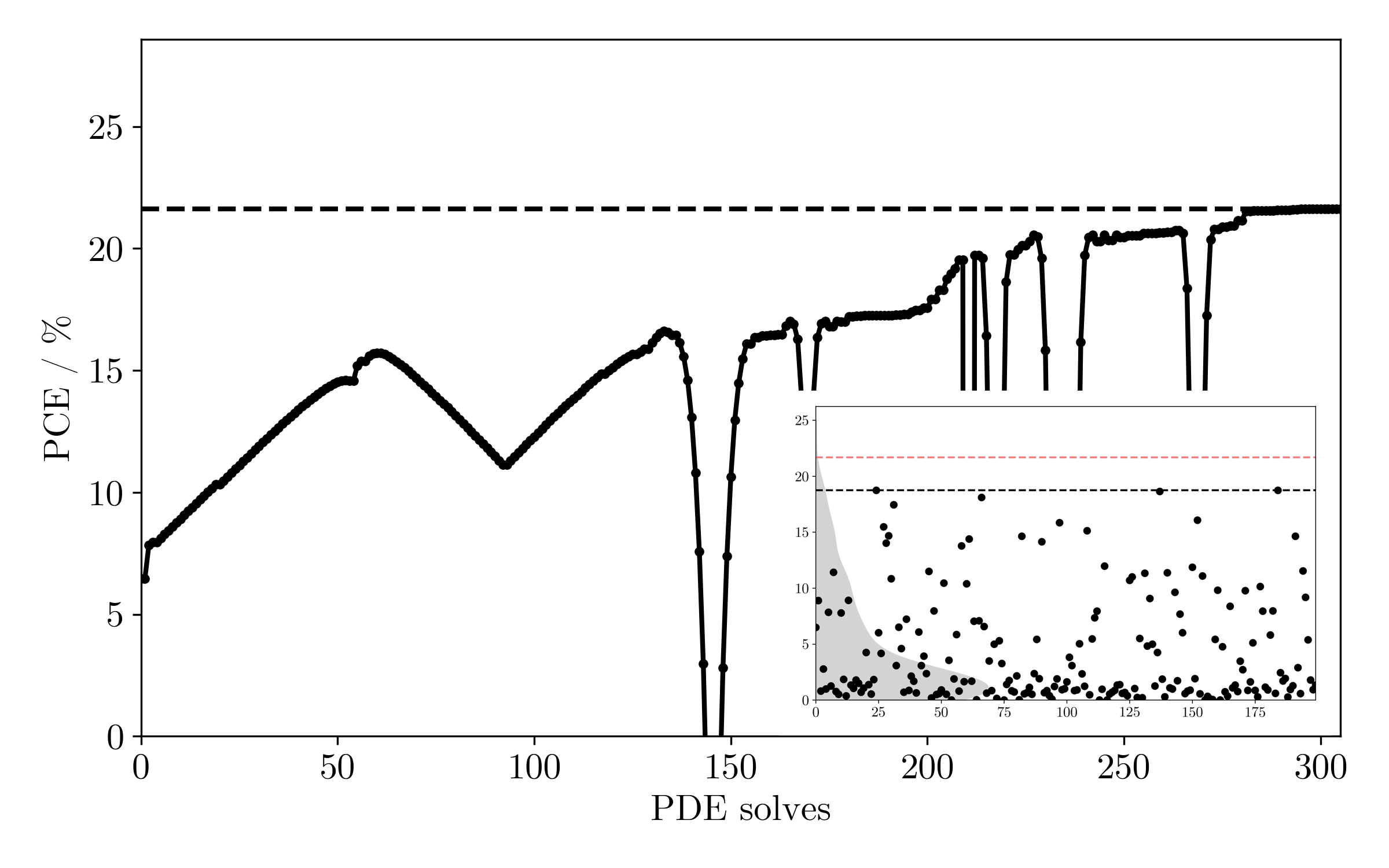}
    \caption{}
    \label{fig:psc_objective}
  \end{subfigure}

  \begin{subfigure}[t]{0.8\textwidth}
    \centering
    \includegraphics[width=\linewidth]{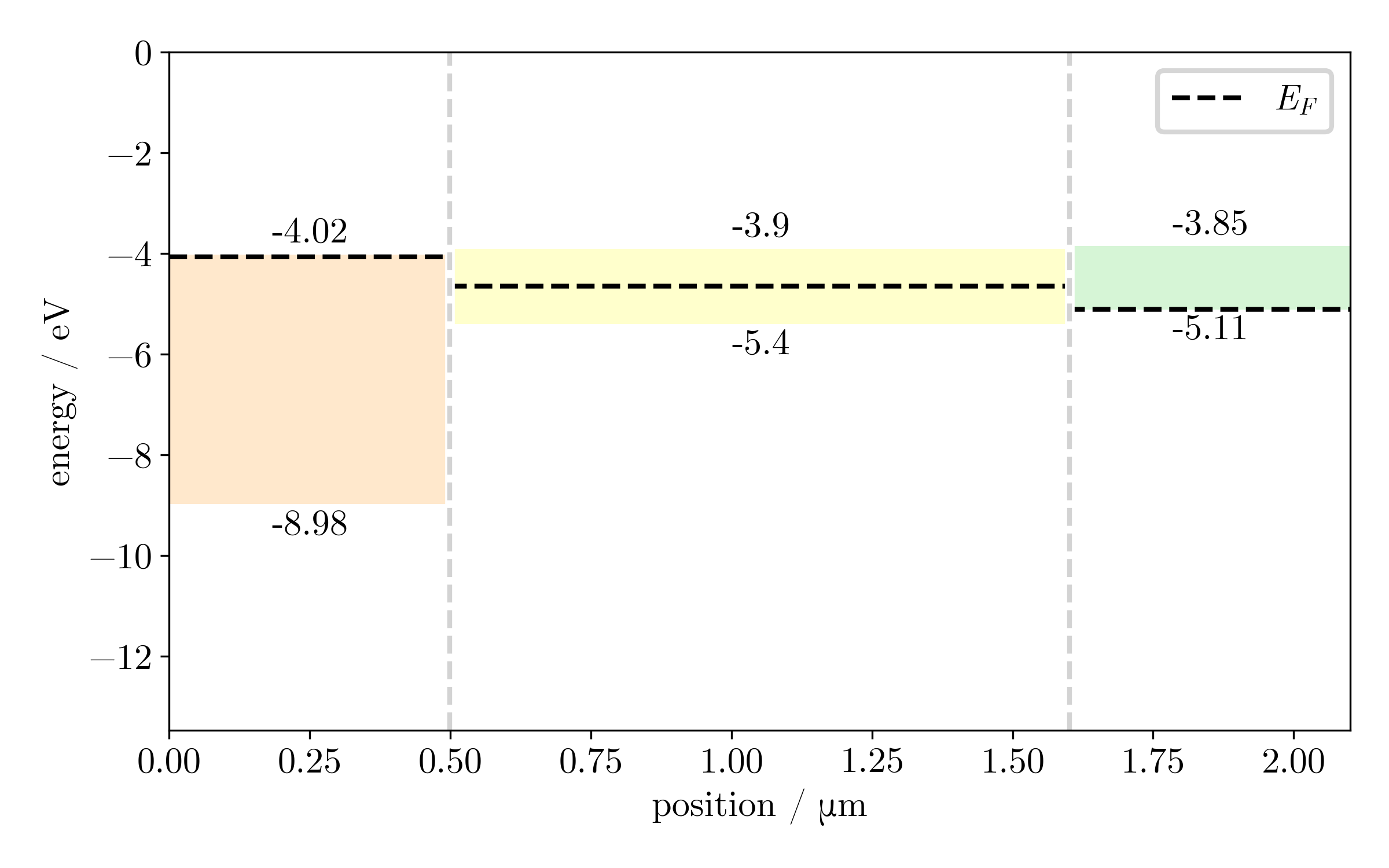}
    \caption{}
    \label{fig:psc_des}
  \end{subfigure}
  
  \caption{Results of using SLSQP to optimize the ETL and HTL material properties of a PSC. a) Growth of the optimization objective compared with random search. The inset contains the PCE of 200 randomly sampled cells together with a kernel-estimated distribution, with the SLSQP optimal point (red) being superior to all samples while being more than 10 times more efficient in terms of PDE solves. Scattered points represent random samples. b) Isolated energy levels of the optimized PSC design.}
  \label{fig:pscoptim}
\end{figure}

To illustrate this, we take the case of optimizing a p--i--n perovskite cell, which has been a device attracting much recent interest in the PV community~\cite{perov}. In particular, we optimize the band gap $E_g$, electron affinity $\chi$, dielectric constant $\epsilon$, electron and hole densities of states (DOS) $N_c, N_v$ and mobilities $m_n, m_p$, and dopant densities $N_a, N_d$ for both the electron (ETL) and hole transport layers (HTL). This amounts to 16 parameters. In addition, we employ reasonable box constraints and the following band-alignment constraints taken from a previous work \cite{pvoptim}:

\begin{equation}
\begin{split}
  \chi_{ETM} - \Phi_{F} &\leq 0 \\
  \chi_{ETM} - \chi_P &\leq 0 \\
  \Phi_{B} - \chi_{HTM} - E_{g, HTM} &\leq 0 \\
  \chi_{HTM} + E_{g, HTM} - \chi_P - E_{g, P} &\leq 0 \\
  \chi_P - \chi_{ETM} &\leq 0
\end{split}
\end{equation}
where $\Phi_F, \Phi_B$ are the front and back contact workfunctions. For simplicity, we compute them via the flat-band approximation by equating the work functions to the Fermi energies at $x = 0, L$. Denoting the intrinsic Fermi energy of a material as $E_i$, we have the following nonlinear relationships

\begin{equation}
    \begin{split}
    \Phi_F &= E_{i, 0} + k_B T \log \left(\frac{N_{d, 0}}{n_{i, 0}}\right) \\
    \Phi_B &= E_{i, L} - k_B T \log \left(\frac{N_{a, L}}{n_{i, L}}\right).
\end{split}
\end{equation}
For our nonlinear constrained optimization problem we select the Sequential Least Squares Programming (SLSQP) method~\cite{kraft1989slsqp}. SLSQP has been implemented in several open-source tools, including Optim~\cite{mogensen2018optim}, NLOpt~\cite{johnson2014nlopt}, PyOpt~\cite{perez2012pyopt} and Scipy~\cite{virtanen2020scipy}. For this work, we chose the last one. Starting from a randomly sampled initial design with 6.49\% PCE, the algorithm terminates after only 306 PDE solves, arriving at an optimal point with PCE 21.62\%. This is higher than that of all 200 randomly sampled designs, which amount to roughly 4000 PDE solves. This is illustrated in Figure \ref{fig:psc_objective}.

As mentioned above, in \dpv it is possible to treat $V$ as an additional variable to be differentiated through via the IFT as part of the intermediate parameters $\tilde{\mathbf{p}}$. At an optimal point, $V$ will be selected as the corresponding MPP of the associated cell. Optimizing $V$ and $\mathbf{p}$ holistically rather than separately brings computational benefits, especially when two consecutive iterations have similar device parameters; in fact, we only need to sweep from these two close points in the $\tilde{\mathbf{p}}$ space rather than sweeping $V$ twice from $0$. In this case, when changing $V$ of about 0.05 V (with all the other parameters varied accordingly), we obtain an overall savings in PDE solves of about $25\%$. The alternative procedure of optimizing $V$ and $\mathbf{p}$ separately is described in~\ref{diff}.

\subsection{Materials Discovery}

The usefulness of gradients is not limited to direct optimization purposes. Here we present a simple demonstration of the potential of an adjoint model built within an AD framework. Unlike the previous example, here we need to compute the whole $IV$ curve, therefore we only consider a subset of $\mathbf{p}$ as optimization parameters. For simplicity, suppose we have an empirical $IV$ curve of a simple p-n homojunction measured from a physical experiment. All material parameters are assumed to be known but for two: the band gap of the material making up the cell, $E_{g}$, and the hole mobility, $\mu_p$. The idea is that given reasonable initial guesses of the parameters, one can recover the latent material property, assuming identifiability of $IV$ curves, by minimizing a dissimilarity measure between the candidate and target $IV$ curves, $\hat{J}$ and $J^*$ respectively. Formally, the problem to solve is to minimize over $\mathbf{p}^*$, the unknown parameters,

\begin{equation}\label{inte}
    R(\hat{J}(\mathbf{p}^*), J^*) := \int_0^{\frac{\pi}{2}} \left[ r^*(\theta) - \hat{r}(\theta, \mathbf{p}^*)\right]^2 d\theta
\end{equation}
where $r^*(\theta)$ refers to the empirical $IV$ curve reparameterized in polar coordinates, with the conventions that $\theta = 0$ corresponds to the positive $y$-axis and $\theta$ grows clockwise. The term $\mathbf{p}^*$ has only the parameters to be discovered (two in this case). The gradient of $\hat{r}$ with respect to $\mathbf{p}^*$ are computed via AD. Eq.~\ref{inte} corresponds to integrating the squared radial differences between the two curves over the first quadrant. The choice of parameterizing by $\theta$ comes from the observation that $IV$ curves are monotonically decreasing from the point $(0, I_{\mathrm{sc}})$ to $ (V_{\mathrm{oc}}, 0)$, and resemble arcs. $IV$ curves are computed over $\left[0, V_{\mathrm{oc}}\right]$, meaning that two curves generally have different support in $V$, causing discontinuities if we were to integrate differences only in the intersection. In contrast, the support of $\left[0, \pi/2\right]$ is always promised in polar coordinates. Meanwhile, the distance measure chosen avoids this issue, and additionally presents pleasant optimization properties, such as smoothness and symmetry, that play a crucial role in the success of this methodology. Note that in a practical implementation, quadratic interpolation and a discrete approximation of the integral are used.

\begin{figure}
 
  \begin{subfigure}[t]{.5\textwidth}
    \centering
    \includegraphics[width=\linewidth]{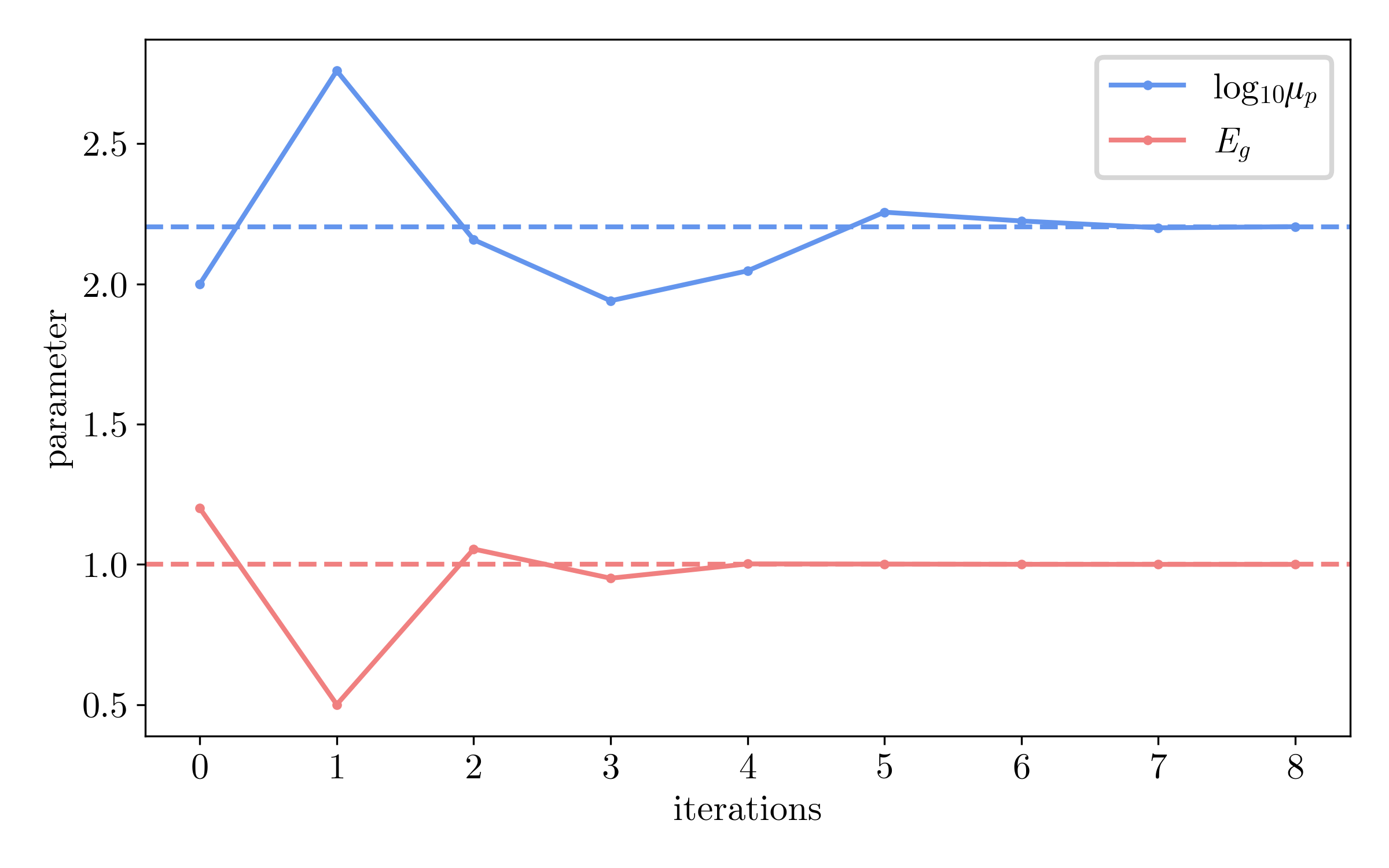}
    \caption{}
  \end{subfigure}
  \hfill
  \begin{subfigure}[t]{.5\textwidth}
    \centering
    \includegraphics[width=\linewidth]{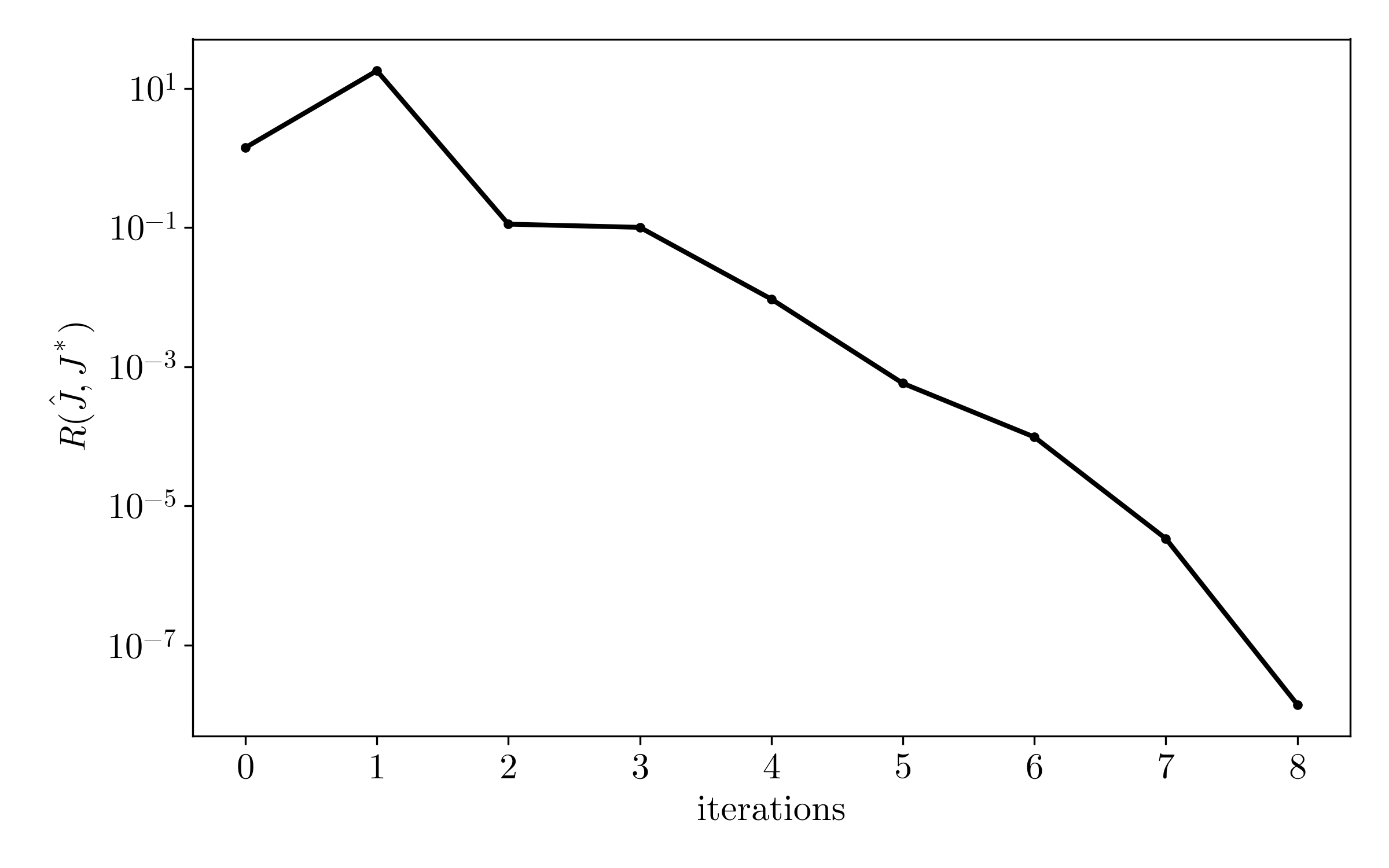}
    \caption{}
  \end{subfigure}
   \caption{SLSQP results for minimizing the distance between a candidate $IV$ curve and the empirical curve. In this minimal example, the process converges within less than 10 function calls. (a) Trajectory of $E_{g}$ and $\log_{10} \mu_p$, the estimated latent parameters. (b) The objective over the descent process in log scale. Note it converges to zero rapidly.}
  \label{fig:discovery}
\end{figure}

SLSQP was once again used for this problem. The algorithm takes less than 10 function calls to arrive at the correct parameters of $E_g = 1.0, \log_{10} \mu_p = 2.20$ from initial values of $1.2$ and $2.0$ respectively. It must be stressed that while the empirical $IV$ curve is generated through simulation for the sake of this example, in a meaningful application it would come from an experiment. (See Fig. ~\ref{fig:discovery})

\section{Conclusion}

We have presented a composable, differentiable PV simulator that, while enabling new data-efficient optimization and materials discovery techniques, does not compromise on computational speed. Being built with sparse linear-algebra compatibility in mind, \dpv has the potential to be easily extended to 2-dimensional or 3-dimensional modeling in the future. This promises more realistic and general modeling for developing novel structures. With JAX also rapidly growing in popularity, the ecosystem for working with \dpv is likely to grow as well.

\section{Acknowledgments}

We thank Rachel Kurchin and Matthias auf Der Maur for helpful discussions, Arthur Reiner De Belen for writing an initial draft for the online documentation, and the referee for the suggestion of adding the section on the sensitivity analysis. This work was supported in part by a fellowship from Eni S.p.A. and the MIT Energy Initiative. We are also grateful to the MIT Energy Initiative and MIT Quest for Intelligence for providing support through the UROP program.

\appendix\section{The Drift-Diffusion model}
\label{physics}

\begin{table}
  \centering
  \begin{tabular}{|c|p{0.3\textwidth}|c|p{0.3\textwidth}|}
  \hline
  \multicolumn{4}{|c|}{Constants} \\
  \hline
  Notation
  & Physical Quantity
  & Notation
  & Physical Quantity \\
  \hline
  $\epsilon_{0}$
  & vacuum permittivity
  &$q$
  & elementary charge \\
  $k_{B}$
  & Boltzmann constant
  &$T$
  & temperature \\
  \hline
  \multicolumn{4}{|c|}{Material Parameters} \\
  \hline
  Notation
  & Physical Quantity
  & Notation
  & Physical Quantity \\
  \hline
  $\epsilon$ 
  & relative permittivity
  & $E_{g}$
  & band gap \\
  $\chi$
  & electron affinity
  & $N_{c}$
  & density of states at the conduction band \\
  $N_{v}$
  & density of states at the valence band
  & $N_{a}$
  & acceptor dopant density \\
  $N_{d}$
  & donor dopant density
  & $\mu_{n}$
  & electron mobility \\
  $\mu_{p}$
  & hole mobility
  & $B$
  & radiative recombination coefficient \\
  $C_{n}$
  & electron Auger coefficient
  & $C_{p}$
  & hole Auger coefficient \\
  $E_t$
  & trap energy level for SRH recombination
  & $t_{n}$
  & bulk electron lifetime \\
  $t_{p}$
  & bulk hole lifetime
  & $S_{n}$
  & electron surface recombination velocity \\
  $S_{p}$
  & hole surface recombination velocity
  &
  & \\
  \hline
  \multicolumn{4}{|c|}{Variables} \\
  \hline
  Notation
  & Physical Quantity
  & Notation
  & Physical Quantity \\
  \hline
  $\phi$
  & electrostatic potential
  & $\phi_{n}$
  & electron quasi-Fermi energy \\
  $\phi_{p}$
  & hole quasi-Fermi energy
  & $n$
  & electron density \\
  $p$
  & hole density 
  & $J_{n}$
  & electron current \\
  $J_{p}$
  & hole current
  &
  & \\
  \hline
  \end{tabular}
\caption{
Notation for the physical quantities used in the model. In the model described in this work, the temperature is considered constant at $300K$. Variables are physical quantities that must be solved for in the solar cell equations.
}

\label{tab:notations}
\end{table}

The DD method couples the continuity equations for electrons and holes with the Poisson equation. In the absence of a temperature gradient it reads as

\begin{equation}
\begin{split}
\epsilon_{0} \vec{\nabla} \cdot ( \epsilon \vec{\nabla} \phi ) &= q( n - p + N_{a} - N_{d} ) \\
\vec{\nabla} \cdot \vec{J}_{n} &= - q ( G - R ) \\
\vec{\nabla} \cdot \vec{J}_{p} &= q ( G - R ) \\
\vec{J}_{n} &= q \mu_{n} n \vec{\nabla} \phi_{n} \\
\vec{J}_{p} &= q \mu_{p} p \vec{\nabla} \phi_{p} \\
\end{split}
\end{equation}
where $R$ is the total recombination rate density, $G$ is the electron-hole pair generation density, $N_a$ ($N_d$) the acceptor (donor) density, $\epsilon$ the relative dielectric constant, $\mu_{n(p)}$ the electron (hole) mobility, $q$ the electron charge, and $\phi_{n(p)}$ the electrochemical potential for electrons (holes). The electron ($n$)and hole ($p$) densities are given by
\begin{equation}
\begin{split}
n &= N_{c} \exp\left( \frac{ \phi_{n} + \chi + q\phi }{ k_{B}T } \right) \\
p &= N_{v} \exp \left( \frac{ - \phi_{p} - \chi - E_{g} - q\phi }{ k_{B}T } \right) 
\end{split}
\end{equation},
where $\chi$ is the electron affinity, $k_B$ the Boltzmann constant, and $E_g$ the energy gap.

The radiative, Auger and Shockley-Read-Hall (SRH) recombination terms are readily expressed as a function of the electron/hole densities and the intrinsic carrier density $n_i = \sqrt{N_{c}N_{v}}\exp({\frac{E_{g}}{2k_{B}T}})$:

\begin{equation}
\begin{split}
R_{radiative} &= B( np - n_i^{2} ) \\
R_{Auger} &= ( C_{n}n + C_{p}p ) ( np - n_i^{2} ) \\
R_{\mathrm{RSH}} &= \frac{ np - n_i^{2} }{ t_{n}\left( n + n_i\exp( \frac{E_t}{k_{B}T} ) \right) + t_{p} \left( p + n_i\exp(- \frac{E_t}{k_{B}T} ) \right)}, \\
\end{split}
\end{equation}
where $E_t$ is the trap energy level, $C_{n(p)}$ the electron (hole) Auger coefficient, and $t_{n(p)}$ is the electron (hole) bulk lifetime. All the material parameters are listed in Table~\ref{tab:notations}.

\section{Optics}\label{optics}

The calculation of the generation density $G(x)$ requires an absorption model for the incoming phonon flux. To this end, we use the simple Beer--Lambert approximation 
\begin{equation}
  S(\lambda, x) = S_{\mathrm{sun}}(\lambda) \exp \left(-\int_0^x \alpha(\lambda, x') dx'\right),
\end{equation}
where $S(\lambda, x)$ is the spectral irradiance of the incoming radiation at distance $x$ and for wavelength $\lambda$. The direct band gap model is used to calculate the absorption coefficient as follows:
\begin{equation}
  \alpha(\lambda, x) = \begin{dcases}
    A \sqrt{\frac{hc}{\lambda} - E_g(x)} &\text{if } \frac{hc}{\lambda} \geq E_g(x) \\
    0 &\text{otherwise}
  \end{dcases}
\end{equation}
The photon flux density is then given by
\begin{equation}
  \varphi(x) = \int_{0}^{\infty} \frac{S(\lambda, x)}{hc/\lambda} d\lambda
\end{equation}
In order to integrate over the highly ragged solar spectrum, we employ a Gaussian quadrature scheme \cite{quadrature} which computes, in advance, a small number $M$ of nodes $\tilde{\lambda}_i$ and their respective effective weights $\tilde{I}_i$ which together approximate to great accuracy integrals of the form
\begin{equation}
  \int_0^{\infty} S_\text{sun}(\lambda) f(\lambda) d\lambda \approx \sum_{i=1}^{M} \tilde{I}_i f(\tilde{\lambda}_i)
\end{equation}
The generation density can then be found by differentiating the photon flux, giving
\begin{equation}
  G(x) = \sum_{i=1}^{M} \frac{\tilde{\lambda}_i \tilde{I}_i}{hc} \alpha(\tilde{\lambda}_i, x) \exp \left(-\int_0^x \alpha(\tilde{\lambda}_i, x')dx' \right)
\end{equation}

\section{The boundary conditions}\label{boundary}

We consider unidimensional solar cells spanning the \textit{x} axis from $x=0$ to $x=L$.  When the system is at equilibrium (no net flow of carriers), there is no current, therefore the quasi-Fermi energies are constant across the system, and are set to zero : $\phi_{n} = \phi_{p} = 0$. Therefore we must solve only the Poisson equation, where we implement the common Dirichlet boundary conditions for Ohmic and Schottky contacts. For Ohmic contacts, depending on whether the contact at $x=0, L$ is $n$-doped or $p$-doped we have:
\begin{equation}
\phi(x=0,L) = 
\begin{dcases}
- \chi + k_{B}T \ln( \frac{ N_{d} }{ N_{c} } ) &\text{$n$-doped} \\
- \chi - E_{g} - k_{B}T \ln( \frac{ N_{a} }{ N_{v} } ) &\text{$p$-doped} \\
\end{dcases}
\end{equation}
Where the front and back contact workfunctions $\Phi_F, \Phi_B$ are provided, we impose the Schottky boundary conditions:
\begin{equation}
  \begin{split}
  \phi(0) &= -\Phi_F \\
  \phi(L) &= -\Phi_B
  \end{split}
\end{equation}
When the system is out of equilibrium and we impose a bias voltage $V$:
\begin{equation}
\begin{split}
\phi(0) &= \phi_{eq}(0) \\   
\phi(L) &= \phi_{eq}(L) + V \\
J_{n}(0) &= q S_{n,0} ( n(0) - n_{eq}(0) ) \\
J_{n}(L) &= - q S_{n,L} \left( n(L) - n_{eq}(L) \right) \\
J_{p}(0) &= - q S_{p,0} \left( p(0) - p_{eq}(0) \right) \\
J_{p}(L) &= q S_{p,L} \left( p(L) - p_{eq}(L) \right) \\
\end{split}
\end{equation}
where $n_{eq},p_{eq}$ are the equilibrium electron/hole densities, and $S_{n/p,0/L}$ are the surface recombination velocities at the $x=0,L$ contacts for electrons/holes.

\section{The Scharfetter--Gummel method}\label{numerical}

We solve the drift-diffusion model using finite differences.
The system is now defined over a grid $x=x_{i}$ for $1 \leq i \leq N$, where a difference is made between material properties and variables defined over the ``grid points" $x_{i}$ and the ``slabs" between consecutive $x_{i} \rightarrow x_{i+1}$ over which the current densities are defined and computed.
Thus, material parameters and variables defined throughout the system are now vectors : $u(x) \rightarrow \{u(x_{i})\}_{1 \leq i \leq N}$.
For the discretization of the current, we use the Scharfetter-Gummel scheme over the ``slabs" \cite{gummel}, which defines trial functions for the current to evaluate the gradient of the currents that appears in the continuity equations, in order to ensure numerical convergence. In this work, we follow the discretization as outlined in SESAME \cite{sesame}, where the currents are defined as follows:
\begin{equation}
\begin{split}
\Psi_{n,i} &= q\phi_{i} + \chi_{i} + k_{B}T\ln(N_{c,i})\\
\Psi_{p,i} &= q\phi_{i} + \chi_{i} + E_{g,i} - k_{B}T\ln(N_{v,i})\\
J^{i\rightarrow i+1}_{n} &= 
- \frac{q\mu_{n,i}}{x_{i+1}-x_{i}}
\frac{\Psi_{n,i+1}-\Psi_{n,i}}{
\exp({ - \frac{\Psi_{n,i+1}}{k_{B}T} }) - \exp({ - \frac{\Psi_{n,i}}{k_{B}T} })
}
\left(
\exp({ \frac{\phi_{n,i+1}}{k_{B}T} }) - \exp({ \frac{\phi_{n,i}}{k_{B}T} })
\right)
\\ 
J^{i\rightarrow i+1}_{p} &= 
\frac{q\mu_{p,i}}{x_{i+1}-x_{i}}
\frac{\Psi_{p,i+1}-\Psi_{p,i}}{
\exp({ \frac{\Psi_{p,i+1}}{k_{B}T} }) - \exp({ \frac{\Psi_{p,i}}{k_{B}T} })
}
\left(
\exp({ -\frac{\phi_{p,i+1}}{k_{B}T} }) - \exp({ -\frac{\phi_{p,i}}{k_{B}T} })
\right)
\\ 
\end{split}
\end{equation}
Therefore, the discretized gradient of the current is:

\begin{equation}
\begin{split}
\frac{dJ_{n}}{dx}\Bigr|_{i} &= \frac{ J^{i\rightarrow i+1}_{n} - J^{i - 1\rightarrow i}_{n} }{ \frac{ x_{i+1} - x_{i-1} }{ 2 } } \\   
\frac{dJ_{p}}{dx}\Bigr|_{i} &= \frac{ J^{i\rightarrow i+1}_{p} - J^{i - 1\rightarrow i}_{p} }{ \frac{ x_{i+1} - x_{i-1} }{ 2 } }    
\end{split}
\end{equation}

Finally, the discretized Laplacian on the left side of the Poisson equation is simply (where the divergence is taken as a central derivative):

\begin{equation}
\frac{d ( \epsilon \frac{d\phi}{dx} ) }{dx}\Bigr|_{i} = 
\frac{1}{ \frac{ x_{i+1} - x_{i-1} }{ 2 } }
\left(
\frac{\epsilon_{i+1} + \epsilon_{i}}{2}
\frac{\phi_{i+1}-\phi_{i}}{ x_{i+1} - x_{i} }
-
\frac{\epsilon_{i} + \epsilon_{i-1}}{2}
\frac{\phi_{i}-\phi_{i-1}}{ x_{i} - x_{i-1} }
\right)
\end{equation}

Thus, the system of differential equations becomes an equation for the zeros of the $F$ function of $3N$ variables $\{\phi_{n,i},\phi_{p,i},\phi_{i}\}_{1 \leq i \leq N} = \{u_{i}\}_{1 \leq i \leq 3N}$, $F:\mathbb{R}^{3N} \rightarrow \mathbb{R}^{3N}$, where 6 components of $F$ are the discretized boundary conditions, and the other $3(N-2)$ are the discretized continuity and Poisson equations.

\section{Linear solver}\label{solver}

To accomodate possible future extensions to 2D and 3D modelling, \dpv computes and stores the band-diagonal Jacobian in a standard compact format allowing only $O(N)$ memory. The linear solve is done via the Generalized Minimum Residual (GMRES) method \cite{gmres}, provided by JAX, which iteratively finds a minimal-residual vector in a Krylov subspace for a large, sparse, and asymmetric system. An $ILU(0)$ preconditioner \cite{ilu0} for band-diagonal matrices was developed to drastically improve the rate of convergence. The preconditioner is used to approximate the inverse of the Jacobian through performining an incomplete $LU$ factorization in $O(N)$ time that preserves the nonzero structure of the Jacobian, which is then used to operate on vectors through the standard two-step forward and backward substitutions.

\section{The modified Newton algorithm}\label{newton}

To avoid stagnation arising from the traditional Newton algorithm, we employ an element-wise damping scheme \cite{sesame} for the Newton step
\begin{equation}
  \tilde{z}^{(i)}_j = \begin{cases}
    z^{(i)}_j &\text{if } -1 \leq z^{(i)}_j \leq 1 \\
    \text{sgn}(z^{(i)}_j)\ln(1 + (e - 1) \cdot |z^{(i)}_j|) &\text{otherwise}
  \end{cases}
\end{equation}
where $e$ refers to Euler's number. We speculate that this damping scheme works well because of the exponential nature of the drift-diffusion system. As an example, consider performing Newton--Raphson on the function $f(x) = e^x - 1$ with an initial guess $x_0 < 0$ to find its root, which is trivially zero. The unmodified Newton step at any $x$ is given by
\begin{equation}
  z = -\frac{e^x - 1}{e^x} = e^{-x} - 1
\end{equation}
Supposing that $x_0$ is sufficiently far away from zero, the Newton step is then dominated by the term $e^{-x_0}$, which is exponential in the correct step $-x_0$. Using the unmodified Newton step would thus cause a huge overshoot, accompanied by numerical overflow issues. It intuitively makes sense to set a threshold on the step size, above which we take the log for a more accurate estimation of the correct step.

It is also interesting to consider the other case, where $x_0 > 0$ with $x_0$ sufficiently far away from zero. Here, $z$ approaches $-1$ regardless of the distance from the root, which results in slow convergence. This is alleviated in \dpv by an additional ``acceleration'' modification on top of the damping as follows
\begin{equation}
  \delta \mathbf{u}^{(i)} = \tilde{\mathbf{z}}^{(i)} + \max \{ \frac{\tilde{\mathbf{z}}^{(i)} \cdot \tilde{\mathbf{z}}^{(i - 1)}}{|\tilde{\mathbf{z}}^{(i)}||\tilde{\mathbf{z}}^{(i - 1)}|}, 0 \} \cdot \delta \mathbf{u}^{(i - 1)}
\end{equation}
which has the effect of ``accumulating'' previous steps when the new damped Newton step is sufficiently similar to the previous one. This effectively allows successive steps to grow quadratically when required.

To ensure convergence of Newton--Raphson, it is standard practice to use the solution at a bias voltage $V_i$ as the initial guess for the next step $V_{i+1}$. \dpv takes this one step further by performing forward extrapolation, using the solutions for $V_i$ and $V_{i-1}$ to compute a significantly more accurate guess of the solution for $V_{i+1}$. Through fitting a quadratic curve, for each grid point, through the three most recent solutions at equally spaced voltages, we obtain
\begin{equation}
  \tilde{\phi}_j^{i+1} = 3\phi_j^{i} - 3\phi_j^{i-1} + \phi_j^{i-2}
\end{equation}
where $\tilde{\phi}_j^{i+1}$ is the initial guess for the electrostatic potential at grid position $j$ for $V_{i+1}$. This simple trick was found to notably improve both convergence behavior and computation time.

\section{Implicit Function Theorem}\label{ifta}
Essentially, we note that the solution to the drift-diffusion system implicitly depends on $\mathbf{\tilde{p}}$ itself, e.g. $\mathbf{f}(\mathbf{u}^*(\mathbf{\tilde{p}}),\mathbf{\tilde{p}})=0$, where $\mathbf{u}^*$ is the solution of Eq.~\ref{linear}. Hence the total derivative of $\mathbf{f}$ with respect to $\mathbf{\tilde{p}}$ is given by
\begin{equation}\label{expansion}
  \mathbf{f}_{\mathbf{u}} \mathbf{u}_\mathbf{\tilde{p}}^* + \mathbf{f}_\mathbf{\tilde{p}} = 0,
\end{equation}
all the quantities being computed at $\mathbf{u}^*$. Equation~\ref{expansion} leads to
\begin{equation}\label{newsol}
  \mathbf{u}^*_\mathbf{\tilde{p}} = -\mathbf{f}^{-1}_{\mathbf{u}} \mathbf{f}_\mathbf{\tilde{p}};
\end{equation}
which provides a much simpler way to obtain the gradients. Note that $\mathbf{f}^{-1}_{\mathbf{u}}$ is just the Jacobian used in the computation of the Newton step, and is thus available for free. Meanwhile, $\mathbf{f}_{\mathbf{\tilde{p}}}$, which follows directly from the drift-diffusion equations, can be obtained easily with AD. While we have bypassed the need for differentiating through iterative solvers, there are still $M$ linear systems to solve in this approach. In fact, a linear system should be solved for each column of $\mathbf{f}_\mathbf{\tilde{p}}$. These problems can be solved by either using a direct method, while reusing the $LU$ factorization, or with an iterative approach, commonly based on block Krylov subspaces. However, here the problem is simplified by taking advantage of \textit{reverse-mode} AD. In fact, this method is preferred over forward-mode AD for wide Jacobians. Our case is the limiting one where the Jacobian has only one row, i.e. the gradient $dI/d\tilde{\mathbf{p}}$. In reverse-mode AD, programs can be seen as the composition of vector-Jacobian-products (VJP), i.e. $\mathbf{v}^T \mathbf{J}$, or, equivalently, $\mathbf{J}^T\mathbf{v}$, where $\mathbf{v}$ is a generic vector. In relation to Eq.~\ref{newsol}, we seek an expression for $\mathbf{v}^T \mathbf{u}^*_\mathbf{p}$. To this end, we simply multiply both sides of Eq.~\ref{newsol} by $\mathbf{v}^T$
\begin{equation}\label{newsol}
  \mathbf{v}^T\mathbf{u}^*_\mathbf{\tilde{p}} = -\mathbf{v}^T\mathbf{f}^{-1}_{\mathbf{u}} \mathbf{f}_\mathbf{\tilde{p}} = \lambda^T \mathbf{f}_\mathbf{\tilde{p}},
\end{equation}
where $\lambda^T$ is the solution of the linear system
\begin{equation}\label{nn2}
 \mathbf{f}_{\mathbf{u}}^T (\mathbf{u}^*) \lambda = -\mathbf{v};
\end{equation}
We note that Eq.~\ref{nn2} represents the adjoint of the linearized original problem (Eq.~\ref{nn}) at $\mathbf{u} = \mathbf{u}^*$~\cite{margossian2019review}. Because reverse AD involves the transposed Jacobian (and hence solving a transpose of the original linear system), reverse-mode AD is also known as an ``adjoint'' method when it is implemented by hand~\cite{strang2007computational}. In \dpv we simply provide the VJP from Eq.~\ref{newsol} and JAX will then compose it with other JVPs, e.g. those arising from Eq.~\ref{cur}, to perform end-to-end reverse AD.

\section{An alternative optimization approach}\label{diff}

\dpv also provides an optimization approach where $V$ and $p$ are optimized separately. In this case, one simulation yields a sequence of $k$ points on the $IV$ curve, $(V_1, I_1) \dots (V_k, I_k)$ and then $P_{\mathrm{MPP}}$ usually taken as the maximum $V_iI_i$ over this discrete set. In \dpv, to ensure differentiability, we interpolate $P(V)$ with a quadratic spline, then analytically maximize to compute the MPP. The PCE $\eta = Q(\mathbf{I})$ is a function of the $k$ points computed on the $IV$ curve, where $\mathbf{I} = \text{vec}(I_1, \dots, I_k)$ is the vectorization of the current $I(\mathbf{u}, \tilde{\mathbf{p}})$ at different voltages. $Q$ denotes the operation of approximating $P(V)$ with a quadratic spline, then maximizing analytically. The bias voltages are evenly spaced with a step size of 0.05 V, so it is not necessary to consider them as explicit variables. The gradient of the efficiency reads
\begin{equation}\label{efficiency}
    \frac{d\eta}{d\mathbf{p}} = Q_\mathbf{\mathbf{I}} \frac{d\mathbf{I}}{d\mathbf{p}} \\
    = \sum_{j=1}^k Q_{I_j} \frac{dI_j}{d\mathbf{p}},
\end{equation}
where $\mathbf{p}$ includes only the input material parameters. The derivative $Q_\mathbf{\mathbf{I}}$ is obtained with AD to link the gradients for each individual PDE solution to the PCE. Meanwhile, we have
\begin{equation}\label{cur2}
  \frac{dI}{d\mathbf{p}}=I_{\tilde{\mathbf{p}}}\tilde{\mathbf{p}}_{\mathbf{p}}
\end{equation}
where, for simplicity, we dropped the subscript $j$. From now on, we can use the IFT, as detailed in~\ref{ifta}.

In \dpv, to ensure smoothly varying results, we use a quadratic spline to construct a once-differentiable interpolation $\tilde{P}(V)$ of the power-voltage curve given the discrete points evaluated, which is a sequence of $k - 1$ quadratic functions $\tilde{P}_1 \dots \tilde{P}_{k-1}$ with the following constraints

\begin{equation}
  \begin{cases}
    \tilde{P}_i(V_i) = V_i I_i &\text{for } 1 \leq i \leq k - 1 \\
    \tilde{P}_{i}(V_{i+1}) = V_{i+1} I_{i+1} &\text{for } 1 \leq i \leq k - 1 \\
    \tilde{P}'_{i}(V_{i+1}) = \tilde{P}'_{i+1}(V_{i+1}) &\text{for } 1 \leq i \leq k - 2 \\
    \tilde{P}'_1(x) \equiv 0
  \end{cases}
\end{equation}
\begin{figure}
    \centering
    \includegraphics[width=0.65\textwidth]{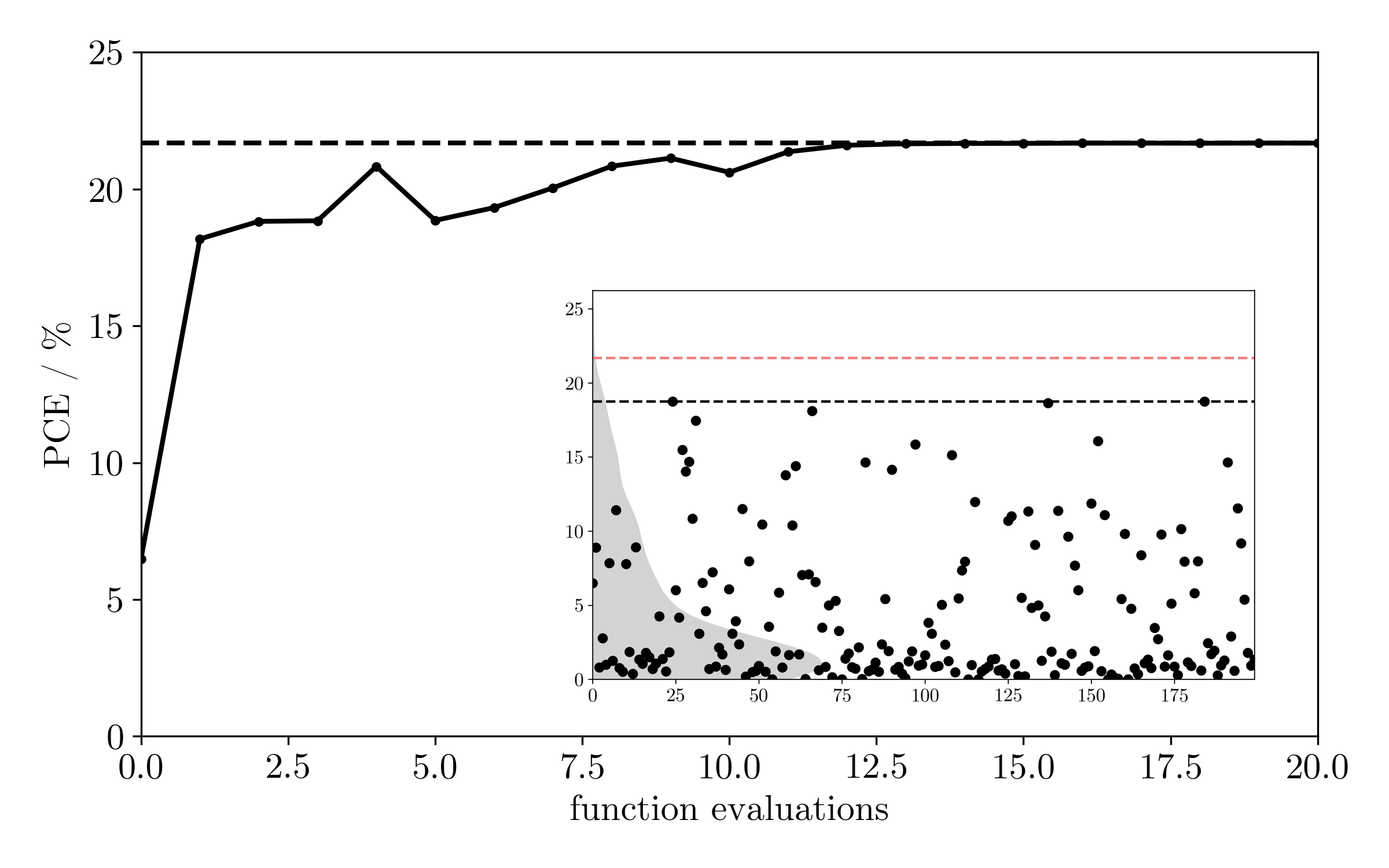}
    \caption{The PCE versus the number of iterations of the SLSQP algorithm. The optimized structure has a PCE of about 21.62$\%$. In the inset, the PCEs for randomnly chosen configurations are shown.\label{opt2}}
\end{figure}
then calculate the MPP analytically by maximizing each $\tilde{P}_i$ over their respective domains. The benefits are two-fold: aside from continuity, this also allows larger voltage steps to be taken while preserving a good estimate of the MPP, and thus the PCE, via spline interpolation. This means that $IV$ curves can typically be computed to a satisfactory degree of precision and accuracy
with only around 20 voltage steps. We use SLSQP algorithm for nonlinear optimization. The final structure has the same PCE as that identified using $V$ as an optimization variable, but with a smoother convergence, shown Fig.~\ref{opt2}. However, the overall number of PDE solves is~400, versus 306 needed for the holistic approach. \dpv provides both optimization techniques.


\bibliographystyle{elsarticle-num}

\bibliography{references.bib}







\end{document}